\documentclass{ws-ijmpd}

\begin{document}

\markboth{J.E. Horvath} {Pulsar astrophysics : the glitch
phenomenon}

\catchline{}{}{}{}{}
%
%%%%%%%%%%%%%%%%%%%%%%%%%%%%%%%%%%%%%%%%%%%%%%%%%%%%%%%%%%%%%%%%%%%%

\title{PULSAR ASTROPHYSICS : THE GLITCH PHENOMENON }

\author{J.E. HORVATH }

\address{ Instituto Astron\^omico e Geof\'\i sico\\
Universidade de S\~ao Paulo\\
R. do Mat\~ao 1226, - Cidade Universit\'aria\\
05508-900 S\~ao Paulo SP, Brazil\\
foton@astro.iag.usp.br}

\maketitle

\begin{history}
\received{Day Month Year} \revised{Day Month Year}
%\accepted{Day Month Year}
%\comby{(xxxxxxxxxx)}
\end{history}

\begin{abstract}
{We briefly discuss a handful of topics in pulsar astrophysics,
first some general well-known features, then an overview of the
glitch phenomenon and the sort of information gathered about the
internal structure and dynamics, and finally the quandary posed by
the precession of PSR B1828-11 a very important clue pointing
towards a novel paradigm for structure of the core regions. We
point out that ``exotic" solutions for the precession puzzle would
force a consideration of exotic {\it glitch} mechanisms as well. }
\end{abstract}

\section{Pulsars and their environments}

Pulsars are now a ``classical" subject of modern astrophysics
after almost four decades of intense work. Shortly after their
discovery in 1967 \cite{HB}, beautiful theoretical work
\cite{Pac,Gold} convincingly argued that neutron stars (and not,
for example, white dwarfs) were responsible for the emission. As a
brand new field at that time, several ideas were put forward and
contributed to fundament the broad-brush picture of pulsars
available today. Thus, concepts such as the charged magnetosphere,
light cylinder and so on form now (in spite of the lack of exact
solutions for this complicated plasma problem), a body of concepts
subject to continuous testing (see Ref.4 for a comprehensive
discussion.

This is not the appropriate place to recall the spectacular
advances in high-resolution instrumentation (see this volume), but
the availability of enhanced ground (Keck, Arecibo, etc.) and
space (HST, Chandra, XMM, etc.) facilities, coupled with the
intensive long-term monitoring (radio) of a handful of objects and
targeted searches have now revealed a wealth of phenomena not
always fitting into the ``standard" view. This has in turn
enriched our vision of pulsars, and also created puzzles for the
models which are being worked out, as is the case of the glitch
phenomenon in particular. We may say that pulsar physics is
definitely entering its maturity where more detailed models can be
constructed and tested.

\section{Torques, braking and glitches : basic picture and challenges}

Pulsars are often depicted as giant rotating dipoles.The simplest
vacuum torque of a magnetized rotating neutron star (with magnetic
dipole $BR^{3}$, angular velocity $\Omega$ and angle $\alpha$
between the magnetic and rotation axis) reads

\begin{equation}
I \Omega {\dot \Omega} = - \frac{2}{3c^{3}} {(BR^{3})^{2}}
\Omega^{4} sin^{2}\alpha
\end{equation}

in spite of the modifications introduced by the currents, it is
expected that the r.h.s. in eq.(1) remains a good representation
of the torque as long as the field remains dipolar (i.e. multipole
contributions are negligible). If so then so-called {\it braking
index} $n={\ddot \Omega}{\Omega}/{\dot \Omega}^{2}$ and {\it jerk
parameter} $m = {d\over{dt}}{\ddot \Omega}\times ({\Omega}/{\dot
\Omega}^{3})$ adopt simple, fixed numerical values (3 and 15
respectively) and deviations may be interpreted as evidence of
varying geometry or non-dipolar character of the emission.

Occasionally, the (otherwise smoothly decreasing) pulsar rotation
frequency $\Omega$ experiences sudden increases ({\it glitches})
and relaxes back to pre-glitch values. The average pulse, on the
other hand, is not observed to change and therefore it is widely
believed that these phenomena reflect the dynamics of the internal
rotating components rather than magnetospheric phenomena. Because
the observed relaxation timescales are very long on microscopic
standards, glitches are currently interpreted as evidence for {\it
superfluid} components, which decouple and recouple as observed
\cite{Ben}.

\begin{figure}[th]
\centerline{\psfig{file=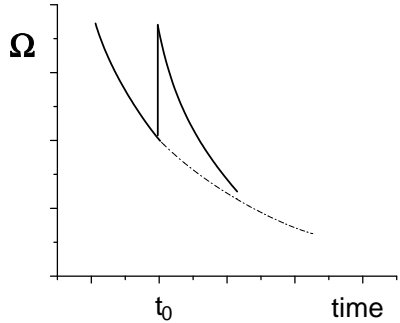,width=6cm}} \vspace*{12pt}
\caption{Schematic draw of a glitch. In this case the angular
rotation frequency relaxes back to its pre-glitch value, as is
often the case, on a variety of increasingly long timescales.}
\end{figure}

As is well-known, the first model devised to explain glitches
invoked cracking of the solid crust (i.e. a starquake) stressed by
the slowdown of the pulsar. It is now believed that these models
are no longer tenable if they are to reproduce the large glitches
observed from the Vela pulsar and a few other objects \cite{Sha},
because not enough elastic energy could be stored in it, although
we shall see below that cracking may still play a role in
glitches.

The most recent approach to a glitch model states that some
interior component displays a variable coupling to the
environment. If glitches actually reflect instead a variable
coupling between the crust (producing the observed pulses through
rigid coupling to the magnetosphere) and a {\it superfluid}
interior (see below), the natural questions to ask are : where
does this coupling occur ? how is the decoupling triggered ? and
how does relaxation proceed ? Each of these questions are closely
related to the issue of the microphysical state of the matter, and
hence to the structure of the neutron star. Because the candidate
superfluids are likely to be located relatively near the surface
(for instance, neutron superfluids are thought to exist between
the neutron drip point and the nuclear saturation density with
neutrons paired in the ${^1}S_{0}$ state), their properties should
be calculable to a high degree of confidence, or at least better
than the supranuclear regime.

One of the issues that has been discussed over the years which
provides a concrete way of addressing these questions is the
possible {\it pinning} of the superfluid vortices to the lattice
of nuclei in the inner crust. This is an ``ideal" place to see
some action (decoupling and recoupling), since calculations
\cite{Pin} suggest that vortices are energetically forced to pin
to a site in the lattice (with energy differences $\leq \, 1 \,
MeV$) at least for a static structure. Because of the rotation
slowdown, torques brake the crust and a velocity difference
develops between the lattice and the superfluid. The vortices
actually {\it creep} radially outwards through the lattice in
steady state. However,this is a gentle collective motion, and
therefore can not be responsible for the sudden hiccup of the
crust shown in Fig. 1 . What is needed to explain the observations
is a {\it sudden} motion of the vortices away from the rotation
axis. The basic picture of pinned vortices is depicted in Fig.2.

\begin{figure}[th]
\centerline{\psfig{file=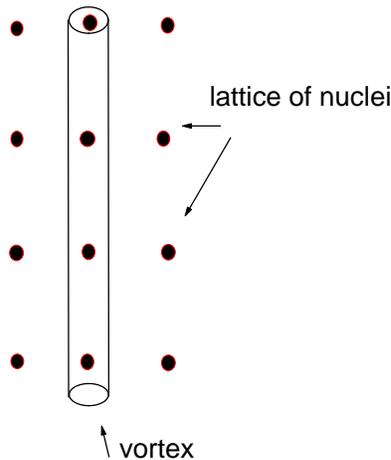,width=6cm}} \vspace*{12pt}
\caption{Pinning of superfluid vortices to the nuclear lattice.
Quantum mechanics requires vortex formation in a neutron
superfluid coexisting with a lattice of neutron-rich nuclei, and
these vortices minimize the energy by pinning to the lattice
sites. The exact realization of the pinning (interstitial,
multiple, etc.) is being debated, but some form of pinning is
required by modern glitch models, differing otherwise in several
important details.}
\end{figure}

If pinned vortices are responsible for the glitch behavior (not
taking into account the starquake model in its original form) two
broad classes of glitch mechanisms, may be constructed to provide
the sudden motion of them. The {\it mechanical models} postulate
that vortices unpin catastrophically (for example, because of a
critical threshold of the velocity difference between the lattice
and the superfluid). The {\it thermal models} in turn search for a
big perturbation of the vortex creep process (for example, because
of an energy deposition). Even though the models are constructed
using the available knowledge of the microphysics (but see below),
they happen to have different relaxation after the glitch and a
few other features. Thus, careful observations can in principle
discriminate between the two.

Since the thermal models require a trigger to perturb the steady
motion of the vortices, it is interesting to note that the old
idea of starquakes has gained a new role as such. The key feature
is that an amount of energy in the form of heat

\begin{equation}
E_{heat} \, \propto \, \mu \theta^{2} \leq 10^{42} erg
\end{equation}

is released per quake event (where $\mu$ is the bulk modulus of
the crust and $\theta$ the critical strain for the fracture to
occur). Since the vortex creep is exponentially sensitive to the
temperature, a starquake trigger turns it into a highly
dissipative motion. A prediction is that $\Omega$ should rise
slower than the mechanical models and relax following the behavior
of the local temperature $T(t)$.

What do observations tell us ? A general overview may be found in
Lyne, Shemar and Graham Smith \cite{LSG}. In the framework of
vortex pinning theory,the huge difference in the glitch behavior
of the Vela and Crab pulsars suggest that the force per unit
length exerted by the vortices on the lattice is different by many
orders of magnitude, thus requiring either a very different
structure in both objects, or rather suggesting again a whole
reinterpretation of the glitch phenomenon (see, for example, Ref.
8 for an argument of this kind). On the other hand, work performed
by Larson and Link \cite{LarLin}showed that fitting of actual
glitches is possible within both the mechanical and thermal
models, although (strangely again) the former seem preferred for
the Vela events and the latter for the Crab. This again may be
indicative of some fundamental flaw in the models if one believes
that a single underlying mechanism should be operating. An
alternative would be to put the blame on evolutionary causes for
these differences, as done, for example, in Ref.10.

\section{More trouble with glitches: quick jumps and ``anomalous" behavior}

The glitch characterization and understanding may seem complicated
enough according to the above remarks. However, accurate
observations continue to reveal a great richness of the glitch
phenomenon, still searching for a firmly established paradigm. A
recent example of that observations can be found in the work by
Dodson, McCulloch and Lewis \cite{trio}, which report accurate
observations of the largest Vela pulsar glitch, fully accelerating
in less than 40 s and relaxing on a series of timescales with a
very short one of $\sim \, 1 \, min$. Even more puzzling than the
short relaxation timescale is the report of the {\it lack} of
relaxation in some of the Crab events, pointing to an increase of
the external torque or an extremely long $\sim \, years$
recoupling of a fraction of the decoupled components in the
standard interpretation (see Fig. 3).

Working within the varying torque hypothesis, models of a growing
angle between the magnetic and rotation axis (\cite{Allen,LiEp})
have been published. Even the growth of the magnetic field
intensity $B$ as suggested earlier \cite{Curt}, although on longer
timescales is in principle possible. Related consequences were
worked out, most notably specific predictions for the
non-canonical braking indices of a small group of selected
pulsars. this quantity is directly measurable and picks up extra
terms which cause deviations from the pure dipole value when the
torque increases after a glitch. The present observational
situation is unclear to us, but there may be indications of a
complete relaxation back to the pre-glitch values on very long
timescales. Nevertheless, it is certain that a successful model of
glitches must have a built-in explanation for very long relaxation
timescales and very short ones, preferably supported by detailed
microphysical calculations.

\begin{figure}[th]
\centerline{\psfig{file=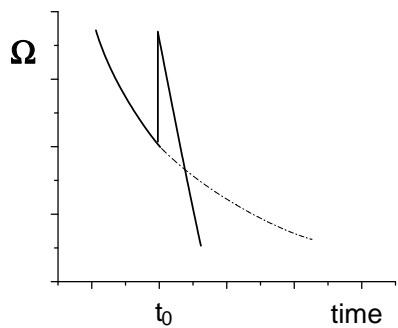,width=6cm}}
\vspace*{12pt}
\caption{The ``anomalous" glitches of the Crab
pulsar. In these events the pulsar is observed to relax only
partially to its pre-glitch state. Since the torque is $\propto \,
{\dot{\Omega}}$, which increases after all the observed
relaxation, one possible explanation is that the geometry of the
field or the field itself have changed. Even if the relaxation is
complete after several years, models would have to explain why
some component remains decoupled for such a long time}
\end{figure}

\section{Precession vs. vortices : type I superfluids or exotic stars ?}

Given that a complex dynamical behavior is present in the data,
and of course that we would like to know more about neutron star
interiors as a whole, it is important to seek for other evidence
to obtai further clues. In a recent paper Link \cite{Li}has argued
that the evidence for precession from PSR B1828-11 is incompatible
with the current models of the {\it outer core} of a compact star.
He showed that the interaction of flux tubes (permeating the
charged superconductor) with rotational vortices (threading the
neutron superfluid) would damp out the precession quickly and
allow high frequency motion only, not one with $\tau_{p} \simeq \,
1 yr$ as suggested by the existing data. There are a few ways out
from this problem, one is that the outer core is actually a type I
superconductor, and therefore expulsion of $B$ by the Meissner
effect happens. Another is that the superfluid and superconductor
do not coexist anywhere in the core. A third possibility, already
raised in that work \cite{Li}, is that an exotic core occurs, the
one we would like to comment on here.

Even though it might appear as if Link's argument may find a
natural realization in the already existing ``hybrid" structure
models (i.e. compact stars with quark cores), the actual situation
is much worse than that: since the existence of nuclear matter in
the outer core would quickly damp the precession motion, its place
in the star must then be taken by the exotic core. However, this
is quite difficult to achieve, because then the transition at
which the phase change starts has to be tuned to be $\simeq
\rho_{0}$. Needless to say, this is far too low to be fashionable.
To quantify the difficulty it is enough to note that in the models
of hybrid stars with CFL cores constructed by Alford and Reddy
\cite{AR} the central density is above $\rho_{0}$ for stellar
masses $M \leq 0.4 \, M_{\odot}$. The ``exotic core" solution can
then be restated in a strong form : unless the nuclear saturation
density is the true threshold value for quark matter to appear,
the quark region is rather likely to extend all the way up to zero
pressure, (i.e. it ``naturally" corresponds to a self-bound state
like strange matter \cite{Wit} or color-flavor-locked strange
matter \cite{LugH} ). Based on this observation we contend that,
if exotic cores as needed to justify the existence of precession
are present, then the {\it locus} of observed glitches must also
be ``exotic" (i.e. unrelated to neutron vortex arrays), simply
because the normal inner crust would not exist. This ``exotic"
glitching models are a promising arena of research for the next
future. ``Ancient" models involving differentiated structures in
quark matter (a prerequisite for any successful attempt to model
glitches) \cite{BHV,HH} have not been explored because of a
disbelief in the employed physics, but are revived in different
new forms from time to time \cite{Xu} and may be worth the effort.
Of course, it is also possible to find out a non-exotic solution
within neutron physics, although it would bring striking novelties
in itself for pulsar structure almost by definition. So stay tuned
for the precession news!.

\section{Conclusions}

We have given a broad overview of one of the most spectacular
dynamical features of pulsars (glitches), repeatedly associated
with perhaps the most gigantic quantum fluids found in nature,
namely the components of the crusts of neutron stars. Accurate
timing for over three decades have provided very good data still
waiting for a comprehensive explanation. Strong limits to the idea
of heat release in a glitch have been set recently by Helfand et
al. \cite{Helf} and Pavlov et al. \cite{tocayo} using Chandra data
to show that the temperature of the pulsar did not change by more
than $0.2 \%$ one month after the Vela glitch of January 2000.
Even though it is not impossible that some mechanism can get rid
of the heat very quickly, these observations constrain the thermal
models in which a large energy input is needed. An even more
serious challenge has been posed by some authors (notably Jones ,
\cite{PBJ} and references therein) suggesting that vortices in the
crust do {\it not} pin at all (see also Donati and Pizzochero
\cite{Pizz} for a general analysis). May be the core plays a role
\cite{JM}, but as discussed above, this component surely hides
some (big ?) surprises and would require extensive studies. We are
far from a thorough understanding of the body of evidence of
glitches, whereas additional complications from related
observations have enriched the general picture recently. A whole
new synthesis is needed soon, and perhaps a change in the paradigm
as well to pin down the essentials of pulsar dynamics.

\section*{Acknowledgements}

The author wishes to thank G. Lugones and M.P. Allen in
particular, and the participants of the Workshop for sharing their
expertise on these topics with him. A. Pereyra is warmly
acknowledge for his scientific advise. Financial support of the
S\~ao Paulo State Agency FAPESP through grants and fellowships,
and the partial support of the CNPq (Brazil) enabled the present
work.

\end{document}